\documentclass{article}

\usepackage[preprint]{neurips_2024}

\usepackage[utf8]{inputenc} 
\usepackage[T1]{fontenc}    
\usepackage{hyperref}       
\usepackage{url}            
\usepackage{booktabs}       
\usepackage{amsfonts}       
\usepackage{nicefrac}       
\usepackage{microtype}      
\usepackage{xcolor}         
\usepackage{algorithm}%
\usepackage[noend]{algorithmic}%
\usepackage{listings}%
\usepackage{adjustbox}%
\usepackage{amsmath}
\usepackage{booktabs} 
\usepackage{array} 
\usepackage{multirow}
\DeclareUnicodeCharacter{202F}{ }
\title{Compressing Recurrent Neural Networks for FPGA-accelerated Implementation in Fluorescence Lifetime Imaging}

\author{
  Ismail Erbas\thanks{Equal contribution.} \\
  Center for Modeling, Simulation, \& Imaging in Medicine\\
  Rensselaer Polytechnic Institute\\
  Troy, NY 12206 \\
  \texttt{erbasi@rpi.edu} \\
  \And
  Vikas Pandey\footnotemark[1] \\
  Center for Modeling, Simulation, \& Imaging in Medicine\\
  Rensselaer Polytechnic Institute \\
   Troy, NY \\
  \texttt{pandev2@rpi.edu} \\
  \And
  Aporva Amarnath \\
  IBM T.J. Watson Research Center \\
  Yorktown Heights, NY  \\
  \texttt{aporva.amarnath@ibm.com} \\
  \AND
  Naigang Wang \\
  IBM T. J. Watson Research Center \\
  Yorktown Heights, NY \\
  \texttt{nwang@us.ibm.com} \\
  \And
  Karthik Swaminathan \\
  IBM T.J. Watson Research Center \\
  Yorktown Heights, NY  \\
  \texttt{kvswamin@us.ibm.com} \\
   \And
  Stefan T. Radev \\
  Center for Modeling, Simulation, \& Imaging in Medicine\\
  Rensselaer Polytechnic Institute \\
  Troy, NY \\
  \texttt{radevs@rpi.edu} \\
   \And
  Xavier Intes \\
  Center for Modeling, Simulation, \& Imaging in Medicine\\
  Rensselaer Polytechnic Institute \\
   Troy, NY \\
  \texttt{intesx@rpi.edu} \\
}

\begin{document}

\maketitle

\begin{abstract}
Fluorescence lifetime imaging (FLI) is an important technique for studying cellular environments and molecular interactions, but its real-time application is limited by slow data acquisition, which requires capturing large time-resolved images and complex post-processing using iterative fitting algorithms. Deep learning (DL) models enable real-time inference, but can be computationally demanding due to complex architectures and large matrix operations.
This makes DL models ill-suited for direct implementation on field-programmable gate array (FPGA)-based camera hardware. Model compression is thus crucial for practical deployment for real-time inference generation. In this work, we focus on compressing recurrent neural networks (RNNs), which are well-suited for FLI time-series data processing, to enable deployment on resource-constrained FPGA boards. We perform an empirical evaluation of various compression techniques, including weight reduction, knowledge distillation (KD), post-training quantization (PTQ), and quantization-aware training (QAT), to reduce model size and computational load while preserving inference accuracy. Our compressed RNN model, Seq2SeqLite, achieves a balance between computational efficiency and prediction accuracy, particularly at 8-bit precision. By applying KD, the model parameter size was reduced by 98\% while retaining performance, making it suitable for concurrent real-time FLI analysis on FPGA during data capture. This work represents a big step towards integrating hardware-accelerated real-time FLI analysis for fast biological processes.
\end{abstract}

\section{Introduction}
Non-invasive imaging methods that reliably capture dynamic cellular processes are key for disease detection, treatment monitoring, and therapy development \cite{dmitriev2021luminescence}. Fluorescence lifetime imaging (FLI) has emerged as a powerful tool for this purpose, providing detailed insights into cellular and molecular activities by measuring fluorescence decay times. Unlike traditional intensity-based fluorescence imaging, FLI is unaffected by factors such as fluorophore concentration or excitation light intensity \cite{mieog2022fundamentals}, making it suitable for investigating complex biological phenomena, including protein-protein interactions and ligand-target binding \cite{dmitriev2021luminescence} in intact small animals.

To date, estimating fluorescence lifetime parameters from time-resolved data remains a computationally intensive task. Conventional approaches rely on post-processing methods after time-resolved data capture, which itself is highly time-consuming. As a result, FLI is not well-suited for applications requiring rapid inference, such as real-time monitoring of fast biological processes or fluorescence-guided surgery. Real-time processing is critical in these contexts, since immediate decisions are necessary. Deep learning models, particularly sequence-to-sequence (Seq2Seq) architectures based on Gated Recurrent Units (GRUs) \cite{pandey2024temporal}, have shown promise in addressing the computational bottlenecks of FLI data analysis. 
These models are well-suited for handling time-series data and can provide faster and more efficient deconvolution of temporal point spread function (TPSF) signals. However, deploying recurrent models on hardware-constrained platforms, such as FPGAs, comes with unique implementational challenges due to limited memory and computational resources.

To address these challenges, the current paper focuses on compressing GRU-based Seq2Seq models for real-time FLI data processing on FPGAs by reducing memory usage and computational complexity.
We explore quantization techniques for weight reduction, such as PTQ \cite{frantar2022optimal,maly2023simple,wang2022deep} and QAT \cite{hubara2018quantized, govorkova2022autoencoders}, to lower the precision of model precision, improving efficiency without a significant drop in accuracy. 
Additionally, we implement knowledge distillation \cite[KD;][]{shin2020knowledge,polino2018model,mishra2018apprentice} to further compress the model by transferring knowledge from a larger model to a smaller one without compromising performance.

\begin{figure*}[t]
\centering\includegraphics[width=13cm]{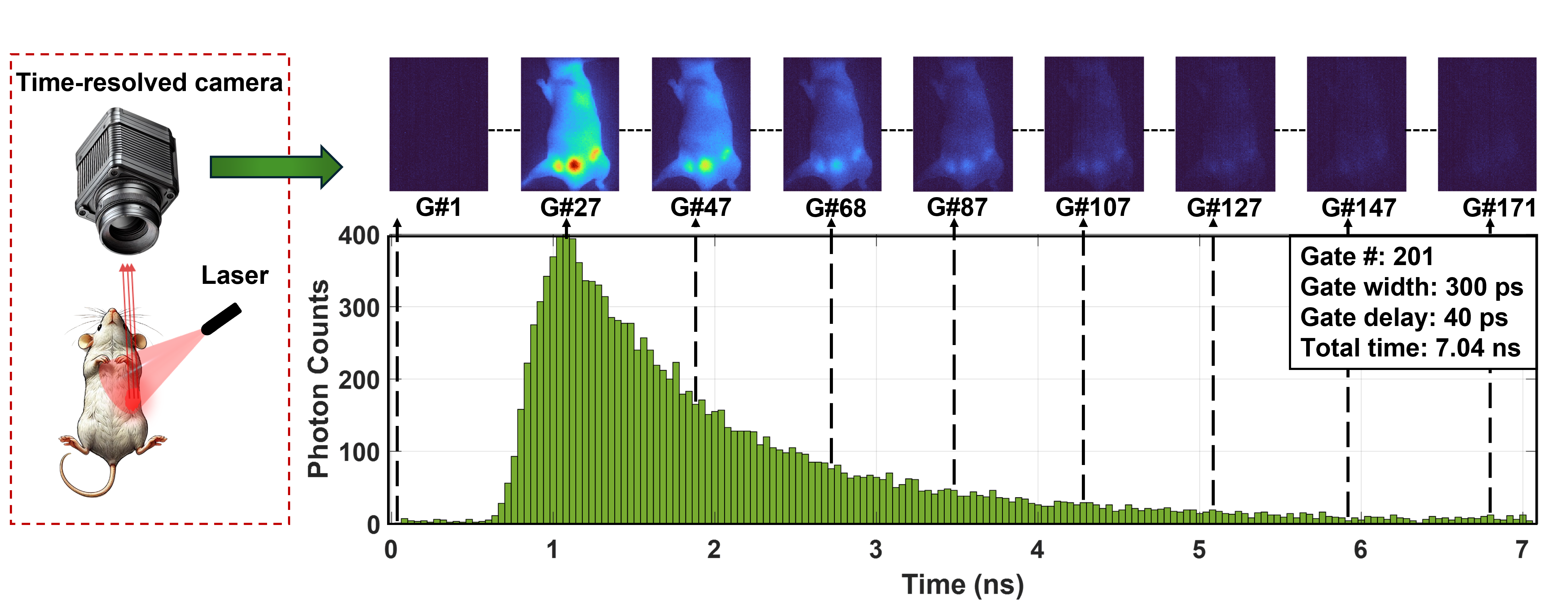}
\caption{\textbf{Experimental set-up and data}. Schematic illustration of fluorescence lifetime imaging (FLI), time-resolved data capture, and the temporal point spread function (TPSF). The top-right panel shows experimental time-resolved fluorescence images of HER2+ tumor xenografts labeled with Alexa Fluor 700 conjugated to Trastuzumab in a nude mouse, captured at different time gates. The bottom-right panel presents the corresponding TPSF for a single pixel.}
\label{fig:tpsf}
\end{figure*}

\section{Background}
FLI operates through time-resolved imaging, where a fluorescent sample is excited with a short pulse of light and the emitted fluorescence signal is captured over time using time-resolved/time-gated detectors such as time-correlated single photon counting (TCSPC), time-gated instensified charge-couple device (ICCD), and single photon avalanche diode (SPAD) \cite{bruschini2019single}. This process generates the TPSF, representing photon-arrival temporal distribution after the excitation pulse (see \autoref{fig:tpsf}). The TPSF is distorted by the Instrument Response Function (IRF), representing the temporal response of the imaging system on the delta input signal. Hence, the observed TPSF can be modeled as a convolution of the sample fluorescence decay (SFD) and the IRF \cite{chen2019vitro}. In biological samples, multiple fluorophore components or changes in local environmental conditions can alter the fluorescence decay rate. These variations contribute to the SFD, which is modeled as the sum of the exponential decay of each fluorophore component.

Despite the advantages of FLI, accurately extracting SFD is computationally expensive. Traditional approaches, such as nonlinear least-squares fitting, center-of-mass, and maximum likelihood estimation methods, require significant computational resources and depend on initial guess parameters\cite{becker2012fluorescence}. These iterative computation methods require IRF estimation prior to the re-convolution and fitting approach.
Hence, they are less practical for fast FLI parameter estimation; DL methods are slowly replacing these computational methods\cite{smith2019fast, erbas2024fluorescence}. Recently, recurrent neural networks (RNNs), specifically GRU networks \cite{cho2014learning}, have demonstrated potential in deconvolving SFD from the TPSF without the need of IRF, allowing for faster and more accurate fluorescence lifetime parameter extraction \cite{pandey2024temporal}. GRUs excel at handling sequential time-series data and reduce the computational complexity compared to traditional methods, making real-time analysis feasible.

This deconvolution method \cite{pandey2024temporal}, integrated in the camera FPGA, can be used for simultaneous temporal data capture and deconvolution operation, which can be a step towards real-time FLI parameter analysis. Such a hardware-accelerated camera will be suitable for clinical diagnostics and dynamic biological process monitoring, where rapid, accurate processing of FLI data is important. However, integration of this DL model architecture into FPGA is highly challenging due to its size and complexities of computation operation. One promising approach, as demonstrated by \cite{lin2024coupling}, involves integrating RNNs into SPAD camera systems to improve the speed and accuracy of FLI data processing. The proposed method bypasses conventional histogram-based techniques by directly estimating fluorescence lifetimes from raw photon timestamps using GRU and LSTM models. While the system, using a $32 \times 32$ pixel SPAD camera, achieved real-time processing with up to 4 million photons per second and frame rates of 10 frames per second, the lower resolution of the camera limited the overall data volume.


\begin{figure*}[t]
\centering\includegraphics[width=13cm]{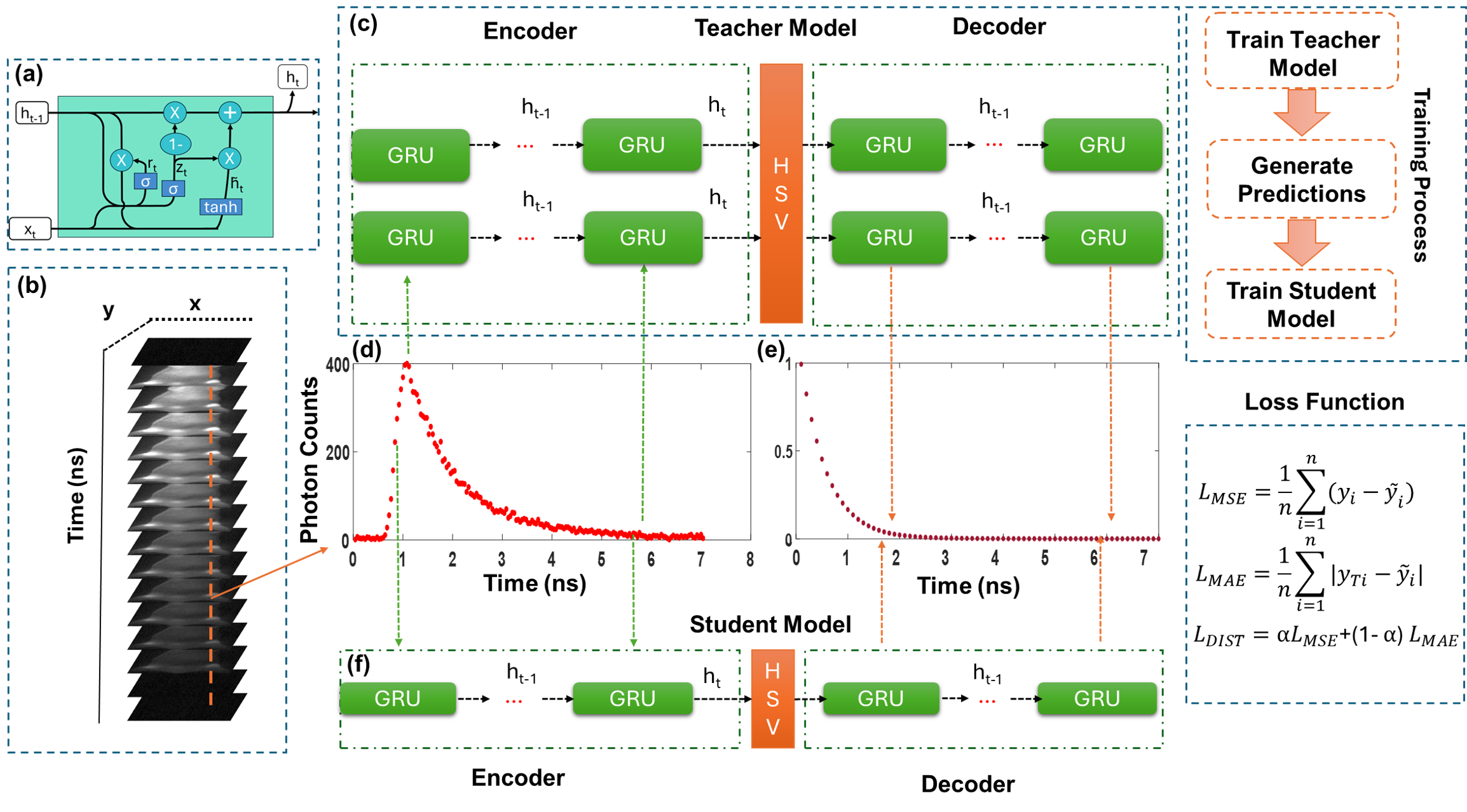}
\caption{\textbf{Model and training setup}. (a) Gated Recurrent Unit (GRU); (b) Time-resolved fluorescence images; (c) Deep GRU-based encoder-decoder architecture (teacher), trained for TPSF deconvolution to pixel-wise SFD from (b), with the resulting deconvolved SFDs shown in (e); (f) Single-layer encoder-decoder RNN model (Student), derived from (c) using the knowledge distillation (KD) method. The stack of time-resolved fluorescence images (b) and deconvolved SFDs (e) are used to train (f). The student model learns hidden features using a combined loss function.}
\label{fig:model}
\end{figure*}

Deploying deep learning model on large-format time resolved detector such as the SwissSPAD2 (SS2) \cite{bruschini2019single} introduces new challenges due to the increased data volume, which places additional demands on FPGA for memory and computation. The SS2 camera with its large-format 512 x 512 time-gated SPAD detector arrays, sub-10-cps dark count rate (DCR) per pixel and 50\% maximum photon detection probability (PDP) has been shown \cite{smith2022vitro} most suitable for fast FLI applications. 

To address these challenges and constraints, DL model compression techniques including quantization methods become highly important \cite{han2015deep}. Quantization reduces the precision of model parameters from floating-point to lower-precision fixed-point formats. This reduces the memory footprint and computational demand, making the models more suitable for FPGA deployment \cite{sun2020ultra}. Quantization can be performed through two main approaches: PTQ and QAT. PTQ involves applying quantization to a pre-trained model without further training, offering a quick but sometimes less accurate solution \cite{frantar2022optimal,maly2023simple,wang2022deep}. In contrast, QAT incorporates quantization during the training process, allowing the model to adapt to reduced precision and maintain higher accuracy, albeit at the cost of additional training time \cite{hubara2018quantized,govorkova2022autoencoders}.
Another key model compression strategy is knowledge distillation \cite{shin2020knowledge,polino2018model,mishra2018apprentice}, where a smaller, more computation efficient model (student) is trained to replicate the performance of a larger, more complex model (teacher). The student model learns from the original training data as well as teacher model’s output, that is helpful to capture key data features while reducing computational requirements. This approach is particularly valuable for deploying deep learning models on hardware-constrained platforms like FPGAs, where maintaining accuracy while reducing model size is critical for FLI applications.

In this work, we explore optimization techniques such as quantization and weight size reduction for the deployment of efficient FLI deconvolution on large-format time resolved detector using an FPGA board. By optimizing these processes, we aim to enhance the performance of real-time FLI in resource-constrained environments, broadening its potential applications in both biomedical research and clinical diagnostics.

\section{Methods}
\subsection{Synthetic data}
To train and validate GRU-based model for FLI, we generated synthetic data simulating time-resolved SFDs. Fluorescence decays were modeled using a bi-exponential function:

\begin{equation}
f(t) = A_R \exp\left(-\frac{t}{\tau_1}\right) + (1 - A_R) \exp\left(-\frac{t}{\tau_2}\right) + \epsilon_t,
\end{equation}

where $\tau_1 \in [0.2, 0.8]$ and $\tau_2 \in [0.8, 1.5]$ represent short and long lifetime components typically observed in near-infrared (NIR) applications (measured in nanoseconds; ns), respectively, and $A_R \in [0, 1]$ denotes the amplitude fraction.
The residual term $\epsilon_t$ denotes system-generated Poisson-distributed  noise.
We used the MNIST dataset to create $28 \times 28$ pixel images with simulated fluorescence decays assigned to each pixel.
To mimic the experimental conditions, we convolved these decays with pixel-wise instrument response functions (IRFs) obtained by illuminating a diffused white paper with a 700 nm laser and capturing the reflected light through a neutral density filter. A total of 200 synthetic image sets were generated, yielding 3,920,000 TPSFs for training.

\subsection{Experimental data}
For experimental validation, we imaged a $10\,\mu M$ solution of Alexa Fluor 700 (AF700) dye in PBS using the large-format time resolved detector. AF700 is a mono-exponential NIR dye with excitation and emission maxima at $702$ nm and $723$ nm, respectively, and a fluorescence lifetime of approximately $1$ ns in PBS. The dye was excited at $700$ nm, and emission was collected using a $740 \pm 10$ nm filter. The collected TPSFs were used to evaluate the model's performance on real-world data.
\subsection{Model development and optimization}

\paragraph{Seq2Seq model} We implemented the GRU-based sequence-to-sequence (Seq2Seq) architecture from \cite{pandey2024deep} for fast estimation of SFDs. \autoref{fig:model}
The encoder consists of two GRU layers, each with 128 hidden units. The input to the encoder is the TPSF sequence for each pixel. Each GRU cell sequentially processes an input time point $x_t$ and the hidden state $h_{t-1}$, producing an output $y_t$ and an updated hidden state $h_t$.  The decoder mirrors the encoder architecture, generating the output SFD sequences from the encoded representation. A final linear dense layer refines the output. The model was trained using the Adam optimizer with a learning rate of 0.001, with mixed loss function. All the models were trained using Windows 11 system with Intel i9-13900K CPU and Nvidia RTX 4090.

\paragraph{Weight reduction and model quantization} To enable deployment on resource-constrained hardware, we explored weight reduction and quantization techniques to reduce the model size and computational complexity.
We experimented with various configurations of hidden units in the GRU layers, including sizes of $128$, $64$, $45$, $32$, and $16$. This resulted in weight matrices of sizes $128 \times 128$, $128 \times 64$, $64 \times 32$, $64 \times 16$, $45 \times 45$, $32 \times 32$, and $16 \times 16$. We trained and tested these models to find the optimal settings within a reasonable error margin.
For model quantization, we applied PTQ, reducing the precision of the weights from 32-bit floating-point to 16-bit and 8-bit integers. The quantization process is defined as
$q = \text{round} \left( x / s\right)$, where $x$ is the floating-point weight, $q$ is the quantized integer value, and $s$ is the scale factor determined by the range of the floating-point weights, typically calculated as $s = \max \left( |x| \right) / \left( 2^b - 1 \right)$ where $b$ represents the bit-width of the target precision (e.g., 8-bit or 16-bit). This approach ensures the quantized values maintain the highest possible precision within the given range.

\paragraph{Knowledge distillation} To further reduce model complexity, we developed a simplified model called Seq2SeqLite, which consists of a single GRU layer in both the encoder and decoder. We experimented with hidden unit sizes of $128$, $64$, $32$, and $16$. During the training process, QAT was applied to the Seq2SeqLite model. QAT incorporates quantization effects by simulating reduced-precision arithmetic during training, allowing the model to adapt effectively to quantization. This approach helps maintain accuracy after quantization to 16-bit or 8-bit precision.
In addition to QAT, we also employed KD to compensate for the reduced model capacity. In this setup, the Seq2Seq model serves as the teacher, while Seq2SeqLite acts as the student. The student model was trained to minimize a combined loss function as shown in \autoref{fig:model}.

\section{Empirical Evaluation}
All models were evaluated using RMSE, $\text{R}^2$ score, L2 norm, and DTW distance. Together, these metrics provide a comprehensive assessment of model performance in terms of accuracy, error magnitude, and temporal consistency.
All models were trained on the simulated data and evaluated on the experimental well plate data, aggregating the metrics across all pixels.
\subsection{Weight reduction results with Seq2Seq model}
\begin{table}[t]
\centering
\setlength{\tabcolsep}{10pt} 
\renewcommand{\arraystretch}{1.2} 
\scriptsize
\begin{tabular}{@{}lcccccc@{}}
\toprule
\textbf{Seq2Seq model size} & ($\bf{64 \times 64}$) & ($\bf{64 \times 32}$) & ($\bf{64 \times 16}$) & ($\bf{45 \times 45}$) & ($\bf{32 \times 32}$) & ($\bf{16 \times 16}$) \\
\midrule
\textbf{RMSE} & 0.071±0.01 & 0.084±0.01 & \textbf{0.032±0.001} & 0.063±0.01 & 0.1±0.01 & 0.1±0.01\\
\textbf{$R^2$ Score} & 0.86±0.02 & 0.81±0.03 & \textbf{0.96±0.01} & 0.89±0.02 & 0.74±0.04 & 0.76±0.03 \\
\textbf{L2 norm} & 0.60±0.05 & 0.71±0.05 & \textbf{0.32±0.04} & 0.53±0.05 & 0.83±0.05 & 0.79±0.05 \\
\textbf{DTW distance} & 0.49±0.01 & 0.47±0.03 & \textbf{0.35±0.02} & 0.46±0.02 & 0.55±0.04 & 0.48±0.02 \\
\bottomrule
\end{tabular}
\vspace{0.5em}
\caption{Performance metrics for different Seq2Seq model configurations on experimental data.}
\label{tab:seq2seq_wr}
\end{table}
The results of the weight reduction experiments for the Seq2Seq model using 32-bit floating point precision are shown in the \autoref{tab:seq2seq_wr}, illustrating the performance across various sizes, including ($128 \times 64$), ($64 \times 64$), ($64 \times 32$), ($64 \times 16$), ($45 \times 45$), ($32 \times 32$), and ($16 \times 16$). RMSE values range from 0.0374 to 0.0995, with the ($64 \times 16$) model showing the lowest error. The $R^2$ score is highest at 0.96 for the ($64 \times 16$) configuration, while more compressed models like ($32 \times 32$) and ($16 \times 16$) show lower performance. The L2 norm is smallest with the ($64 \times 16$) model, whereas compressed models exhibit larger norms. DTW distance is lowest for the ($64 \times 16$) model and highest for the ($32 \times 32$) model.  Overall, these results highlight the trade-offs between model size and performance, with the ($64 \times 16$) configuration consistently achieving superior results across most metrics. 
 
\subsection{Quantization results}
\begin{table}[h!]
\centering
\setlength{\tabcolsep}{10pt} 
\renewcommand{\arraystretch}{1.2} 
\scriptsize
\begin{tabular}{@{}llcccccc@{}}
\toprule
\multicolumn{8}{c}{\textbf{Seq2Seq Models (16-bit and 8-bit)}} \\
\midrule
\textbf{Type} & \textbf{Metrics} & {$\bf{64 \times 64}$} & {$\bf{64 \times 32}$} & {$\bf{64 \times 16}$} &{$\bf{45 \times 45}$} & {$\bf{32 \times 32}$} & {$\bf{16 \times 16}$} \\
\midrule
\multirow{2}{*}{\textbf{16-bit}} & \textbf{RMSE} & 0.071±0.01 & 0.084±0.01 & \textbf{0.032±0.001} & 0.063±0.01 & 0.1±0.01 & 0.095±0.01 \\
 & \textbf{$R^2$ Score} & 0.87±0.02 & 0.81±0.03 & \textbf{0.96±0.01} & 0.89±0.02 & 0.74±0.04 & 0.76±0.03 \\
 & \textbf{L2 norm} & 0.59±0.05 & 0.71±0.05 & \textbf{0.31±0.04} & 0.53±0.05 & 0.83±0.05 & 0.79±0.05 \\
 & \textbf{DTW distance} & 0.49±0.01 & 0.47±0.03 & \textbf{0.34±0.02} & 0.46±0.02 & 0.55±0.04 & 0.48±0.02 \\
\midrule
\multirow{2}{*}{\textbf{8-bit}} & \textbf{RMSE} & 0.063±0.01 & 0.089±0.01 & \textbf{0.032±0.001} & 0.071±0.01 & 0.105±0.01 & 0.089±0.01 \\
 & \textbf{$R^2$ Score} & 0.89±0.02 & 0.79±0.03 & \textbf{0.97±0.01} & 0.87±0.02 & 0.70±0.04 & 0.79±0.03 \\
 & \textbf{L2 norm} & 0.54±0.05 & 0.74±0.05 & \textbf{0.27±0.04} & 0.58±0.05 & 0.88±0.05 & 0.75±0.05 \\
 & \textbf{DTW distance} & 0.69±0.02 & 0.72±0.04 & \textbf{0.52±0.03} & 0.53±0.02 & 0.61±0.04 & 0.44±0.02 \\
\bottomrule
\end{tabular}
\caption{Performance metrics for Seq2Seq models (16-bit and 8-bit) across various configurations on experimental data.}
\label{tab:seq2seq_combined}
\end{table}

The performance of different Seq2Seq model architectures under PTQ with 16-bit and 8-bit precision was evaluated as shown in \autoref{tab:seq2seq_combined}. The models assessed include configurations such as ($64 \times 64$), ($64 \times 32$), ($64 \times 16$), ($45 \times 45$), ($32 \times 32$), and ($16 \times 16$). For the 16-bit models, the ($64 \times 16$) architecture achieved the lowest RMSE of 0.032, along with an $R^2$ score of 0.96 and a relatively low L2 norm of 0.31, indicating efficient weight scaling. The model also exhibited a low DTW distance of 0.34, making it suitable for applications where both accuracy and computational efficiency are required. In contrast, the ($32 \times 32$) model, while showing reasonable performance with an $R^2$ score of 0.74, demonstrated higher RMSE and L2 norm values, indicating larger errors and weight magnitudes. In the 8-bit setting, the ($64 \times 16$) model continued to perform well, with an RMSE of 0.032 and an $R^2$ score of 0.97. The L2 norm decreased slightly to 0.27, suggesting that the 8-bit quantization did not introduce significant degradation in performance. While DTW distance increased to 0.52, the model maintained competitive performance compared to other configurations. Overall, 8-bit quantized models showed a slight increase in RMSE and DTW distance compared to their 16-bit counterparts, but still performed competitively. The ($64 \times 16$) configuration consistently delivered strong results across both precision levels, making it a favorable choice for memory-constrained deployment scenarios.

The results for the quantized Seq2SeqLite models, focused on memory-efficient architectures, are summarized in \autoref{tab:seq2seqlite_combined}, comparing different configurations under 16-bit and 8-bit quantization, both with and without KD. 
For the 16-bit models, the ($32 \times 32$) architecture with KD achieves the lowest RMSE (0.068), while the ($16 \times 16$) model shows moderate improvement with KD, reducing its RMSE from 0.088 to 0.072. $R^2$ scores indicate better fit when KD is applied, with the ($32 \times 32$) model improving from 0.73 without KD to 0.88 with it, and similar trends are observed for the ($16 \times 16$) architecture. Metrics such as L2 norm and DTW distance also improve with KD.
\begin{table}[t]
\centering
\setlength{\tabcolsep}{10pt} 
\renewcommand{\arraystretch}{1.2} 
\scriptsize
\begin{tabular}{@{}llccccc@{}}
\toprule
\multicolumn{7}{c}{\textbf{Seq2SeqLite Models (16-bit and 8-bit)}} \\
\midrule
\textbf{Type} & \textbf{Metrics} & \textbf{$\bf{128 \times 128}$} & \textbf{$\bf{32 \times 32}$} & \textbf{$\bf{32 \times 32}$ w/ KD} & \textbf{$\bf{16 \times 16}$} & \textbf{$\bf{16 \times 16}$ w/ KD} \\
\midrule
\multirow{2}{*}{\textbf{16-bit}} &\textbf{RMSE} & 0.03±0.000 & 0.1±0.01 & \textbf{0.068±0.01} & 0.088±0.01 & 0.072±0.01 \\
 & \textbf{$R^2$ Score} & 0.98±0.01 & 0.73±0.39 & \textbf{0.88±0.29} & 0.79±0.04 & 0.86±0.03 \\
 & \textbf{L2 norm} & 0.25±0.04 & 0.84±0.05 & \textbf{0.57±0.06} & 0.74±0.06 & 0.60±0.06 \\
 & \textbf{DTW distance} & 0.37±0.03 & 0.50±0.04 & \textbf{0.44±0.04} & 0.49±0.03 & 0.45±0.04 \\
\midrule
\multirow{2}{*}{\textbf{8-bit}} & \textbf{RMSE} & 0.12±0.01 & 0.138±0.02 & \textbf{0.01±0.001} & 0.13±0.01 & 0.047±0.01 \\
 & \textbf{$R^2$ Score} & 0.62±0.08 & 0.49±0.11 & \textbf{0.99±0.01} & 0.55±0.08 & 0.94±0.02 \\
 & \textbf{L2 norm} & 1.00±0.10 & 1.15±0.12 & \textbf{0.10±0.02} & 1.09±0.09 & 0.39±0.06 \\
 & \textbf{DTW distance} & 0.61±0.07 & 0.71±0.08 & \textbf{0.52±0.05} & 0.71±0.06 & 0.55±0.03 \\
\bottomrule
\end{tabular}
\vspace{0.5em}
\caption{Performance metrics for Seq2SeqLite quantized models (16-bit and 8-bit) with and without KD on experimental data.}
\label{tab:seq2seqlite_combined}
\end{table}
In the 8-bit setting, the impact of KD is more pronounced. The ($32 \times 32$) model with KD achieves an RMSE of 0.01, a significant improvement over the non-KD configuration (0.138). Similarly, the ($16 \times 16$) model shows improved performance with KD, reducing RMSE from 0.13 to 0.047. $R^2$ scores and L2 norms confirm the overall benefit of KD across the architectures.

In summary, applying KD improves performance across all architectures in both 16-bit and 8-bit quantized models, particularly in terms of the RMSE, $R^2$, L2 norm, and DTW metrics, making them more suitable for deployment in memory-constrained environments.
\section{Conclusion}

In this work, we compressed DL models for real-time FLI data processing on resource-constrained hardware, such as FPGAs. 
By focusing on GRU-based Seq2Seq architecture, we explored model weight reduction and quantization techniques to enable efficient deployment of these models in FPGAs. Specifically, we applied PTQ and QAT to reduce the precision of model parameters to 16-bit and 8-bit formats. 
Furthermore, we incorporated KD as a model compression strategy to reduce computational complexity while maintaining high performance. The ($32 \times 32$) Seq2SeqLite model with KD demonstrated an optimal balance between model size and performance in both 16-bit and 8-bit quantized versions. The 8-bit version, in particular, showed strong suitability for real-time applications, making it ideal for FPGA deployment. These improvements make the ($32 \times 32$) KD model particularly suited for FPGA deployment in clinical and research settings where both memory efficiency and computational speed are critical.

\bibliography{arxiv}



\newpage

\end{document}